\magnification \magstep1
\raggedbottom
\openup 2\jot
\voffset6truemm

\centerline {\bf PEIERLS BRACKETS IN FIELD THEORY}
\vskip 1cm
\noindent
{\it Giuseppe Bimonte, Giampiero Esposito, Giuseppe Marmo,
Cosimo Stornaiolo}
\vskip 1cm
\noindent
Dipartimento di Scienze Fisiche, Universit\`a Federico II,
Complesso Universitario di Monte S. Angelo, Via Cintia, Edificio N',
80126 Napoli, Italy
\vskip 0.3cm
\noindent
Istituto Nazionale di Fisica Nucleare, Sezione di Napoli,
Complesso Universitario di Monte S. Angelo, Via Cintia, Edificio N',
80126 Napoli, Italy
\vskip 1cm
\noindent
{\bf Abstract}. Peierls brackets are part of the space-time approach to
quantum field theory, and provide a Poisson bracket which, being
defined for pairs of observables which are group invariant, is
group invariant by construction. It is therefore well suited for
combining the use of Poisson brackets and the full diffeomorphism
group in general relativity. The present paper provides an
introduction to the topic, with applications to gauge field theory.
\vskip 100cm

\leftline {\bf 1. Introduction}
\vskip 0.3cm
\noindent
Al\-tho\-ugh the Ha\-mil\-to\-ni\-an for\-ma\-li\-sm
pro\-vi\-des a po\-wer\-ful to\-ol for
stu\-dy\-ing ge\-ne\-ral re\-la\-ti\-vi\-ty,$^{1}$
its initial-value problem and
the approach to canonical quantization,$^{2}$ it suffers from
severe drawbacks: the space $+$ time split of $(M,g)$ disagrees
with the aims of general relativity, and the space-time topology
is taken to be $\Sigma \times {\bf R}$, so that the full diffeomorphism
group of $M$ is lost.$^{3,4}$

However, as was shown by DeWitt in the sixties,$^{5}$ it remains
possible to use a Poisson-bracket formalism which preserves the
full invariance properties of the original theory, by relying upon
the work of Peierls.$^{6}$ In our paper, whose aims are pedagogical,
we begin by describing the general framework, assuming that the
reader has been introduced to the DeWitt covariant approach
to quantum field theory.$^{5}$
Let us therefore consider a gauge field theory with classical
action functional $S$ and generators of infinitesimal gauge
transformations denoted by $R_{\; \alpha}^{i}$. The small
disturbances $\delta \varphi^{i}$ are ruled by the invertible
differential operator
$$
F_{ij} \equiv S_{,ij}+\gamma_{ik}R_{\; \alpha}^{k}
{\widetilde \gamma}^{\alpha \beta}\gamma_{jl}R_{\; \beta}^{l},
\eqno (1)
$$
where $\gamma_{ij}$ is a local and symmetric matrix which is
taken to transform like $S_{,ij}$ under group transformations,
and ${\widetilde \gamma}^{\alpha \beta}$ is a local, non-singular,
symmetric matrix which transforms according to the adjoint
representation of the infinite-dimensional invariance group
(hence one gets $R_{i \alpha} \equiv \gamma_{ij}R_{\; \alpha}^{j}$
and $R_{i}^{\; \alpha} \equiv {\widetilde \gamma}^{\alpha \beta}
R_{i \beta}$, respectively). We are interested in advanced and
retarded Green functions $G^{\pm}$ which are left inverses of
$-F$, i.e.
$$
G^{\pm ij}F_{jk}=-\delta_{\; k}^{i}.
\eqno (2)
$$
Furthermore, the form of $F_{ij}$ and arbitrariness of Cauchy data
imply that $G^{\pm}$ are right inverses as well, i.e.
$$
F_{ij}G^{\pm jk}=-\delta_{i}^{\; k}.
\eqno (3)
$$
If symmetry of $F$ is required, one also finds
$$
G^{+ij}=G^{-ji}, \; G^{-ij}=G^{+ji},
\eqno (4)
$$
because in general
$$
G^{\pm ij}-G^{\mp ji}=G^{\pm ik}(F_{kl}-F_{lk})G^{\mp jl}.
\eqno (5)
$$
Thus, the {\it supercommutator function} defined as
$$
{\widetilde G}^{ij} \equiv G^{+ij}-G^{-ij}
\eqno (6)
$$
is antisymmetric in that ${\widetilde G}^{ij}=-
{\widetilde G}^{ji}$.
These properties show that, on defining
$\delta_{A}^{\pm}B \equiv \varepsilon B_{,i}G^{\pm ij}A_{,j}$,
one has, on relabelling dummy indices,
$$
\delta_{A}^{\pm}B
=\varepsilon B_{,j}G^{\pm ji}A_{,i}
=\varepsilon A_{,i}G^{\mp ij}B_{,j}
=\delta_{B}^{\mp}A.
\eqno (7)
$$
These are the {\it reciprocity relations}, which express the idea
that the retarded (resp. advanced) effect of
$A$ on $B$ equals the advanced (resp. retarded) effect of $B$
on $A$. Another cornerstone of the formalism is a relation
involving the Green function $\widehat G$ of the operator
$-{\widehat F}$, having set $R_{k \beta}R_{\; \alpha}^{k}
\equiv {\widehat F}_{\beta \alpha}$; this is
$$
R_{\; \alpha}^{i} \; {\widehat G}^{\pm \alpha \beta} \;
{\widetilde \gamma}_{\beta \delta}
=R_{\; \alpha}^{i} \; {\widehat G}_{\; \; \; \delta}^{\pm \alpha}
=G^{\pm ij} \; \gamma_{jk} \; R_{\; \delta}^{k}
=G^{\pm ij} \; R_{j \delta}.
\eqno (8)
$$
This holds because, {\it for background fields satisfying the field
equations}, one finds that
$$
F_{ik} R_{\; \alpha}^{k}=R_{i}^{\; \beta}R_{k \beta}R_{\; \alpha}^{k}
=R_{i}^{\; \beta}{\widehat F}_{\beta \alpha}.
\eqno (9)
$$
On multiplying this equation on the left by $G^{\pm ji}$ and on the
right by ${\widehat G}^{\pm \alpha \beta}$ one gets
$$
R_{\; \alpha}^{j}{\widehat G}^{\pm \alpha \beta}
=G^{\pm ji}R_{i}^{\; \beta},
\eqno (10)
$$
i.e. the desired formula (8) is proved. Moreover, by virtue of
(4), the transposed equations
$$
{\widehat G}^{\pm \alpha \beta}R_{\; \beta}^{j}
=R_{i}^{\; \alpha}G^{\pm ij}
\eqno (11)
$$
also hold.
We are now in a position to define the Peierls bracket of any two
observables $A$ and $B$. First, we consider the operation
$$
D_{A}B \equiv \lim_{\varepsilon \to 0}\varepsilon^{-1}
\delta_{A}^{-}B,
\eqno (12)
$$
with $D_{B}A$ obtained by interchanging $A$ with $B$ in (12).
The {\it Peierls bracket} of $A$ and $B$ is then defined by
$$
(A,B) \equiv D_{A}B-D_{B}A=\lim_{\varepsilon \to 0}
{1\over \varepsilon}\Bigr[\varepsilon A_{1}G^{+}B_{1}
-\varepsilon A_{1}G^{-}B_{1}\Bigr]
=A_{1}{\widetilde G}B_{1}=A_{,i}{\widetilde G}^{ij}B_{,j},
\eqno (13)
$$
where we have used (7) and (12) to obtain the last expression.
Following DeWitt,$^{7}$ it should be stressed that the Peierls
bracket depends only on the behaviour of infinitesimal disturbances.

In classical mechanics, following Peierls,$^{6}$ we may arrive at the
derivatives in (12) and (13) starting from the action functional
$S \equiv \int L \; d\tau$ and considering the extremals of $S$ and
those of $S + \lambda A$, where $\lambda$ is an infinitesimal
parameter and $A$ any function of the path $\gamma$. Next we
consider solutions of the modified equations as expansions in powers
of $\lambda$, and hence the new set of solutions to first order
reads
$$
\gamma'(\tau)=\gamma(\tau)+\lambda D_{A} \gamma(\tau).
\eqno (14)
$$
This modified solution is required to obey the condition
that, in the distant past, it should be identical with the original
one, i.e.
$$
D_{A}\gamma(\tau) \rightarrow 0 \; {\rm as} \; \tau
\rightarrow -\infty.
\eqno (15)
$$
Similarly to the construction of the above ``retarded'' solution,
we may define an ``advanced'' solution
$$
\gamma''(\tau)=\gamma(\tau)+\lambda {\cal D}_{A} \gamma(\tau),
\eqno (16)
$$
such that
$$
{\cal D}_{A}\gamma(\tau) \rightarrow 0 \; {\rm as} \;
\tau \rightarrow + \infty.
\eqno (17)
$$
>From these modified solutions one can now find $D_{A}\gamma(\tau)$
along the solutions of the un-modified action and therefore,
to first order,
the changes in any other function $B$ of the field variables, and
these are denoted by $D_{A}B$ and $D_{B}A$, respectively.
\vskip 100cm
\leftline {\bf 2. Mathematical Properties of Peierls Brackets}
\vskip 0.3cm
\noindent
We are now aiming to prove that
$(A,B)$ satisfies all properties of a Poisson bracket. The first
two, anti-symmetry and bilinearity, are indeed obvious:
$$
(A,B)=-(B,A),
\eqno (18)
$$
$$
(A,B+C)=(A,B)+(A,C),
\eqno (19)
$$
whereas the proof of the Jacobi identity is not obvious and is
therefore presented in detail. First, by repeated application of
(13) one finds
$$ \eqalignno{
\; & P(A,B,C) \equiv (A,(B,C))+(B,(C,A))+(C,(A,B)) \cr
&= A_{,i}{\widetilde G}^{il}\Bigr(B_{,j}{\widetilde G}^{jk}
C_{,k}\Bigr)_{,l}+B_{,j}{\widetilde G}^{jl}
\Bigr(C_{,k}{\widetilde G}^{ki}A_{,i}\Bigr)_{,l}
+C_{,k}{\widetilde G}^{kl}
\Bigr(A_{,i}{\widetilde G}^{ij}B_{,j}\Bigr)_{,l} \cr
&= A_{,il}B_{,j}C_{,k}\Bigr({\widetilde G}^{ij}
{\widetilde G}^{kl}+{\widetilde G}^{jl}{\widetilde G}^{ki}\Bigr)
+A_{,i}B_{,jl}C_{,k}\Bigr({\widetilde G}^{jk}
{\widetilde G}^{il}+{\widetilde G}^{kl}{\widetilde G}^{ij}\Bigr) \cr
&+ A_{,i}B_{,j}C_{,kl}\Bigr({\widetilde G}^{ki}
{\widetilde G}^{jl}+{\widetilde G}^{il}{\widetilde G}^{jk}\Bigr) \cr
&+ A_{,i}B_{,j}C_{,k}\Bigr({\widetilde G}^{il}
{\widetilde G}_{\; \; \; ,l}^{jk}
+{\widetilde G}^{jl}{\widetilde G}_{\; \; \; ,l}^{ki}
+{\widetilde G}^{kl}{\widetilde G}_{\; \; \; ,l}^{ij}\Bigr).
&(20)\cr}
$$
Now the antisymmetry property of $\widetilde G$, jointly with commutation
of functional derivatives: $T_{,il}=T_{,li}$ for all $T=A,B,C$,
implies that the first three terms on the last equality in (20)
vanish. For example one finds
$$ \eqalignno{
\; &  A_{,il}B_{,j}C_{,k}\Bigr({\widetilde G}^{ij}
{\widetilde G}^{kl}+{\widetilde G}^{jl}{\widetilde G}^{ki}\Bigr)
=A_{,li}B_{,j}C_{,k}\Bigr({\widetilde G}^{lj}
{\widetilde G}^{ki}+{\widetilde G}^{ji}{\widetilde G}^{kl}\Bigr) \cr
&= -A_{,il}B_{,j}C_{,k}\Bigr({\widetilde G}^{jl}
{\widetilde G}^{ki}+{\widetilde G}^{ij}{\widetilde G}^{kl}\Bigr)=0,
&(21)\cr}
$$
and an entirely analogous procedure can be applied to the terms
containing the second functional derivatives $B_{,jl}$ and
$C_{,kl}$. The last term in (20) requires new calculations
because it contains functional derivatives of ${\widetilde G}^{ij}$.
These can be dealt with after taking infinitesimal variations of
Eq. (3), so that
$$
F \; \delta G^{\pm}=-(\delta F)G^{\pm},
\eqno (22)
$$
and hence
$$
G^{\pm}F \delta G^{\pm}=F G^{\pm} \delta G^{\pm}
=-\delta G^{\pm}=-G^{\pm}(\delta F)G^{\pm},
\eqno (23)
$$
i.e.
$$
\delta G^{\pm}=G^{\pm}(\delta F)G^{\pm}.
\eqno (24)
$$
Thus, the desired functional derivatives of advanced and retarded
Green functions read
$$ \eqalignno{
\; & G_{\; \; \; \; \; ,c}^{\pm ij}=G^{\pm ia}F_{ab,c}G^{\pm bj}
=G^{\pm ia}\Bigr(S_{,ab}+R_{a \alpha}R_{b}^{\; \alpha}\Bigr)_{,c}
G^{\pm bj} \cr
&= G^{\pm ia}S_{,abc}G^{\pm bj}+G^{\pm ia}R_{a \alpha,c}
R_{b}^{\; \alpha} \; G^{\pm bj}
+G^{\pm ia}R_{a \alpha} R_{b \; \; ,c}^{\; \alpha}
\; G^{\pm bj}.
&(25)\cr}
$$
In this formula the contractions $R_{b}^{\; \alpha} \; G^{\pm bj}$
and $G^{\pm ia}R_{a \alpha}$ can be re-expressed with the help of
Eqs. (10) and (11), and eventually one gets
$$
G_{\; \; \; \; \; ,c}^{\pm ij}=G^{\pm ia} S_{,abc}G^{\pm bj}
+G^{\pm ia}R_{a \alpha,c}{\widehat G}^{\pm \alpha \beta}
R_{\; \beta}^{j}
+R_{\; \beta}^{i} \; {\widehat G}_{\; \; \; \alpha}^{\pm \beta}
\; R_{b \; \; ,c}^{\; \alpha} \; G^{\pm bj}.
\eqno (26)
$$
By virtue of the group invariance property satisfied by all
physical observables, the second and third term on the right-hand
side of Eq. (26) give vanishing contribution to (20). One is
therefore left with the contributions involving third functional
derivatives of the action. Bearing in mind that $S_{,abc}=
S_{,acb}=S_{,bca}=...$, one can relabel indices summed over, finding
eventually (upon using (4))
$$ \eqalignno{
\; & P(A,B,C)=A_{,i}B_{,j}C_{,k}\Bigr[(G^{+ic}-G^{-ic})
(G^{+ja}G^{+bk}-G^{-ja}G^{-bk}) \cr
&+ (G^{+jc}-G^{-jc})(G^{+ka}G^{+bi}-G^{-ka}G^{-bi}) \cr
&+ (G^{+kc}-G^{-kc})(G^{+ia}G^{+bj}-G^{-ia}G^{-bj})\Bigr]
S_{,abc} \cr
&= A_{,i}B_{,j}C_{,k}\Bigr[(G^{+ia}-G^{-ia})(G^{+jb}G^{-kc}
-G^{-jb}G^{+kc})\cr
&+ (G^{+jb}-G^{-jb})(G^{+kc}G^{-ia}-G^{-kc}G^{+ia})\cr
&+ (G^{+kc}-G^{-kc})(G^{+ia}G^{-jb}-G^{-ia}G^{+jb})
\Bigr]S_{,abc} =0.
&(27)\cr}
$$
This sum vanishes because it involves six pairs of triple products of
Green functions with opposite signs.
The Jacobi identity is therefore fulfilled. Moreover, the fourth
fundamental property of Poisson brackets, i.e.
$$
(A,BC)=(A,B)C+B(A,C)
\eqno (28)
$$
is also satisfied, because
$$
(A,BC)=A_{,i}{\widetilde G}^{ik}(BC)_{,k}
=A_{,i}{\widetilde G}^{ik}B_{,k}C+BA_{,i}{\widetilde G}^{ik}C_{,k}
=(A,B)C+B(A,C).
\eqno (29)
$$
Thus, the Peierls bracket defined in (13) is indeed a Poisson
bracket of physical observables. Equation (28) can be regarded
as a compatibility condition of the Peierls bracket with the product
of physical observables.

It should be stressed that the idea of Peierls$^{6}$ was to introduce
a bracket related directly to the action principle without making
any reference to the Hamiltonian. This implies that even classical
mechanics should be considered as a ``field theory'' in a
zero-dimensional space, having only the time dimension. This means
that one deals with an infinite-dimensional space of paths
$\gamma: {\bf R} \rightarrow Q$, therefore we are dealing with
functional derivatives and distributions even in this situation
where modern standard treatments rely upon $C^{\infty}$ manifolds
and smooth structures. Thus, the present treatment is hiding most
technicalities involving infinite-dimensional manifolds. In finite
dimensions on a smooth manifold, any bracket satisfying (19) and
(28) is associated with first-order bidifferential operators;$^{8,9}$
in this proof it is important that the commutative and associative
product $BC$ is a local product. In any case these brackets at the
classical level could be a starting point to define a
$*$-product in the spirit of non-commutative geometry$^{10}$ or
deformation quantization.$^{11}$
\vskip 0.3cm
\leftline {\bf 3. The most general Peierls bracket}
\vskip 0.3cm
\noindent
The Peierls bracket is a group invariant by construction, being
defined for pairs of observables which are group invariant, and
is invariant under both infinitesimal and finite changes in the
matrices $\gamma_{ij}$ and ${\widetilde \gamma}_{\alpha \beta}$.
DeWitt [5] went on to prove that, even if
independent differential operators $P_{i}^{\; \alpha}$ and
$Q_{i \alpha}$ are introduced such that
$$
F_{ij} \equiv S_{,ij}+P_{i}^{\; \alpha}Q_{j \alpha}, \;
{\widehat F}_{\alpha \beta} \equiv Q_{i \alpha}R_{\; \beta}^{i}, \;
F_{\alpha}^{\; \beta} \equiv R_{\; \alpha}^{i}
P_{i}^{\; \beta},
\eqno (30)
$$
are all non-singular, with unique advanced and retarded Green
functions, the reciprocity theorem expressed by (7) still
holds, and the resulting Peierls bracket is invariant under changes
in the $P_{i}^{\; \alpha}$ and $Q_{i \alpha}$, by virtue of
the identities
$$
Q_{i \alpha}G^{\pm ij}=G_{\; \; \alpha}^{\pm \; \; \beta}
R_{\; \beta}^{j},
\eqno (31)
$$
$$
G^{\pm ij}P_{j}^{\; \beta}=R_{\; \alpha}^{i}
{\widehat G}^{\pm \alpha \beta}.
\eqno (32)
$$
This is proved as follows. The composition of $F_{ik}$ with the
infinitesimal generators of gauge transformations yields
$$
F_{ik}R_{\; \alpha}^{k}=P_{i}^{\; \beta}F_{\beta \alpha},
\eqno (33)
$$
and hence
$$
G^{\pm ji}F_{ik}R_{\; \alpha}^{k}=-R_{\; \alpha}^{j}
=G^{\pm ji}P_{i}^{\; \gamma}F_{\gamma \alpha},
\eqno (34)
$$
which implies
$$
R_{\; \alpha}^{j}G^{\pm \alpha \beta}=-G^{\pm ji}P_{i}^{\; \gamma}
F_{\gamma \alpha}G^{\pm \alpha \beta}
=G^{\pm ji}P_{i}^{\; \beta},
\eqno (35)
$$
i.e. Eq. (32) is obtained. Similarly,
$$
R_{\; \alpha}^{i}F_{ij}=F_{\alpha}^{\; \beta}Q_{j \beta},
\eqno (36)
$$
and hence
$$
G_{\; \; \alpha}^{\pm \; \; \gamma}R_{\; \gamma}^{i}F_{ij}
=-Q_{j \alpha},
\eqno (37)
$$
which implies
$$
Q_{i \alpha}G^{\pm ij}=-G_{\; \; \alpha}^{\pm \; \; \gamma}
R_{\; \gamma}^{k}F_{ki}G^{\pm ij}
=G_{\; \; \alpha}^{\pm \; \; \beta}R_{\; \beta}^{j},
\eqno (38)
$$
i.e. Eq. (31) is obtained. Now we use the first line of Eq. (7)
for $\delta_{A}^{\pm}B$, jointly with Eq. (5), so that
$$
\delta_{A}^{\pm}B-\varepsilon B_{,i}G^{\mp ji}A_{,j}
=\varepsilon B_{,i}R_{\; \gamma}^{i}G^{\pm \gamma \alpha}
Q_{l \alpha}G^{\mp jl}A_{,j}
-\varepsilon B_{,i}P_{l}^{\; \alpha}G^{\pm ik}Q_{k \alpha}
G^{\mp jl}A_{,j}.
\eqno (39)
$$
Since $B$ is an observable by hypothesis, the first term on the right-hand
side of (39) vanishes. Moreover one finds, from (32)
$$
G^{\pm ik}P_{l}^{\; \alpha}Q_{k \alpha}G^{\mp jl}
=G^{\pm il}R_{\; \beta}^{j}G^{\mp \beta \alpha}Q_{l \alpha}.
\eqno (40)
$$
and hence also the second term on the right-hand side of (39) vanishes
($A$ being an observable, for which $R_{\; \beta}^{j}A_{,j}=0$), yielding
eventually the reciprocity relation (7). Moreover, the invariance of the
Peierls bracket under variations of $P_{i \alpha}$ and $Q_{i}^{\; \alpha}$
holds because
$$ \eqalignno{
\; & \delta (\delta_{A}^{\pm}B)=\varepsilon B_{,i}
\delta G^{\pm ij}A_{,j}
=\varepsilon B_{,i}G^{\pm ik}(\delta F_{kl})G^{\pm lj}A_{,j} \cr
&= \varepsilon B_{,i}G^{\pm ik}\Bigr[(\delta P_{k}^{\; \alpha})Q_{l \alpha}
+P_{k}^{\; \alpha}(\delta Q_{l \alpha})\Bigr]G^{\pm lj}A_{,j} \cr
&= \varepsilon B_{,i}G^{\pm ik}(\delta P_{k}^{\; \alpha})Q_{l \alpha}
G^{\pm lj}A_{,j}+\varepsilon B_{,i}G^{\pm ik}P_{k}^{\; \alpha}
(\delta Q_{l \alpha})G^{\pm lj}A_{,j} \cr
&= \varepsilon B_{,i}G^{\pm ik} (\delta P_{k}^{\; \alpha})
G_{\; \; \alpha}^{\pm \; \; \beta}R_{\; \beta}^{j}A_{,j}
+\varepsilon B_{,i}R_{\; \gamma}^{i}G^{\pm \gamma \alpha}
(\delta Q_{l \alpha})G^{\pm lj}A_{,j}=0,
&(41)\cr}
$$
where Eqs. (31) and (32) have been exploited once more.
\vskip 0.3cm
\leftline {\bf 4. Concluding Remarks}
\vskip 0.3cm
\noindent
The Peierls-bracket formalism is
equivalent to the conventional canonical formalism when the
latter exists. The proof can be given starting from
point Lagrangians, as is shown in Ref. 5. Current
applications of Peierls brackets deal with string theory,$^{12,13}$
path integration and decoherence,$^{14}$ supersymmetric
proof of the index theorem,$^{15}$ classical dynamical systems
involving parafermionic and parabosonic dynamical variables,$^{16}$
while for recent literature on covariant approaches to a
canonical formulation of field theories we refer the reader to
the work in Refs. 17--24.

In the infinite-dimensional setting which, strictly, applies also
to classical mechanics, as we stressed at the end of Sec. 2,
we hope to elucidate the relation between a covariant description
of dynamics as obtained from the kernel of the symplectic form,
and a parametrized description of dynamics as obtained from any
Poisson bracket, including the Peierls bracket.
\vskip 10cm
\leftline {\bf Acknowledgements}
\vskip 0.3cm
\noindent
The work of the authors has been partially supported
by PRIN 2000 and PRIN 2002 {\it Sintesi}.
\vskip 0.3cm
\leftline {\bf References}
\vskip 0.3cm
\noindent
\item {1.}
P. A. M. Dirac, {\it Phys. Rev.} {\bf 114}, 924 (1959).
\item {2.}
B. S. DeWitt, {\it Phys. Rev.} {\bf 160}, 1113 (1967).
\item {3.}
C. J. Isham and K. Kuchar, {\it Ann. Phys.} {\bf 164}, 288 (1985).
\item {4.}
C. J. Isham and K. Kuchar, {\it Ann. Phys.} {\bf 164}, 316 (1985).
\item {5.}
B. S. DeWitt, {\it Dynamical Theory of Groups and Fields}
(Gordon \& Breach, New York, 1965).
\item {6}
R. E. Peierls, {\it Proc. R. Soc. Lond.}
{\bf A214}, 143 (1952).
\item {7.}
B. S. DeWitt, {\it Phys. Rev.} {\bf 162}, 1195 (1967).
\item {8.}
J. Grabowski and G. Marmo, {\it J. Phys.} {\bf A34}, 3803 (2001).
\item {9.}
J. Grabowski and G. Marmo, ``Binary Operations in
Classical and Quantum Mechanics'' (MATH-DG 0201089).
\item {10.}
A. Connes, {\it J. Math. Phys.} {\bf 41}, 3832 (2000).
\item {11.}
J. M. Gracia-Bondia, F. Lizzi, G. Marmo and P. Vitale,
JHEP {\bf 0204}, 026 (2002).
\item {12.}
S. R. Das, C. R. Ordonez and M. A. Rubin, {\it Phys. Lett.}
{\bf B195}, 139 (1987).
\item {13.}
C. R. Ordonez and M. A. Rubin, {\it Phys. Lett.} {\bf B216},
117 (1989).
\item {14.}
D. M. Marolf, {\it Ann. Phys. (N.Y.)} {\bf 236}, 392 (1994).
\item {15.}
A. Mostafazadeh, {\it J. Math. Phys.} {\bf 35}, 1095 (1994).
\item {16.}
A. Mostafazadeh, {\it Int. J. Mod. Phys.} {\bf A11}, 2941 (1996).
\item {17.}
G. Marmo, N. Mukunda and J. Samuel, {\it Riv. Nuovo Cimento}
{\bf 6}, 1 (1983).
\item {18.}
G. Barnich, M. Henneaux and C. Schomblond, {\it Phys. Rev.}
{\bf D44}, R939 (1991).
\item {19.}
B. A. Dubrovin, M. Giordano, G. Marmo and A. Simoni,
{\it Int. J. Mod. Phys.} {\bf A8}, 3747 (1993).
\item {20.}
G. Esposito, G. Gionti and C. Stornaiolo, {\it Nuovo Cimento}
{\bf B110}, 1137 (1995).
\item {21.}
H. Ozaki, HEP-TH 0010273.
\item {22.}
I. V. Kanatchikov, {\it Int. J. Theor. Phys.} {\bf 40}, 1121 (2001).
\item {23.}
I. V. Kanatchikov, GR-QC 0012038.
\item {24.}
C. Rovelli, GR-QC 0111037;
C. Rovelli, GR-QC 0202079; C. Rovelli, GR-QC 0207043.

\bye